*

# Efficient and Lightweight In-memory Computing Architecture for Hardware Security

Hala Ajmi, Fakhreddine Zayer, Amira Hadj Fredj, Belgacem Hamdi, Baker Mohammad, *Senior Member, IEEE,* Naoufel Werghi, *Senior Member, IEEE* and Jorge Dias, *Senior Member, IEEE*

*Abstract*— **The paper proposes in-memory computing (IMC) solution for the design and implementation of the Advanced Encryption Standard (AES) based cryptographic algorithm. This research aims at increasing the cyber security of autonomous driverless cars or robotic autonomous vehicles. The memristor (MR) designs are proposed in order to emulate the AES algorithm phases for efficient in-memory processing. The main features of this work are the following: a memristor 4bit state element is developed and used for implementing different arithmetic operations for AES hardware prototype; A pipeline AES design for massive parallelism and compatibility targeting MR integration; An FPGA implementation of AES-IMC based architecture with MR emulator. The AES-IMC outperforms existing architectures in both higher throughput, and energy efficiency. Compared with the conventional AES hardware, AES-IMC shows ~30% power enhancement with comparable throughput. As for state-of-the-art AES based NVM engines, AES-IMC has comparable power dissipation, and ~62% increased throughput. By enabling the cost-effective real-time deployment of the AES, the IMC architecture will prevent unintended accidents with unmanned devices caused by malicious attacks, including hijacking and unauthorized robot control.**

## I. INTRODUCTION

THE widespread use of digital data for storage and communication created a high-risk to information security. With this vast growth of devices especially in IoT applications, data security is becoming of the main constraint to protect confidential information and artificial intelligence (AI) systems[1]. Lightweight cryptography (LC) is a popular topic consist in providing a high-security level for different devices and sensors in the AI/IoT era. Nowadays, multiple domains are supported and getting benefits from LC, such as, internet of things (IoT) and electronic systems [2].

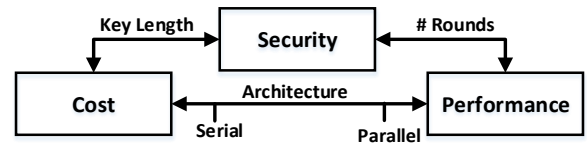

**Fig.1.** Trade-offs among security, performance, and cost.

Encryption algorithms are employed to protect several cryptographic systems due to the large key exchanges and round numbers. However, such algorithms suffer from high power, and computing complexity. Fig.1 shows the trade-off among performance, security and cost for efficient lightweight cryptography. Each system makes a balance between these conflicting requirements to arrive to acceptable point [3]. The Advanced Encryption Standard (AES), for instance, is considered as one of the most popular secure symmetric key cryptography algorithms, due to its low cost and low complexity in both software and hardware designs. This algorithm uses single key to encrypt and decrypt sensitive data[4],[5].

At the technological level, CMOS-based technology is facing many challenges to meet the ever-increasing requirements for computing, communication, and storage especially for the IoTs and edge devices. In addition, von memory-wall issue [6] due to the traditional Von Neuman architecture is adding to the limitation due to the high memory load of many applications such as security and AI. Both Industry and academia, nowadays, are actively developing promising new technologies, and architectures to respond to the new era of AI and big data. On the technology front, emerging non-volatile resistive ram memory (NVM) [7],[8] is good option for low leakage, small size, low active power, CMOS compatibility, and support for crossbar architecture. Crossbar architecture is great option for massive parallelism and efficient implementation of complex computation such as Multiply-add-

*This paragraph of the first footnote will contain the date on which you submitted your paper for review, which is populated by IEEE. It is IEEE style to display support information, including sponsor and financial support acknowledgment, here and not in an acknowledgement section at the end of the article. For example, "This work was supported in part by the U.S. Department of Commerce under Grant BS123456."

*Corresponding author: Fakhreddine Zayer.*
The authors H. Ajmi and F. Zayer contributed equally and are co-first authors.
Hala Ajmi, Amira Hadj Fredj and Belgacem Hamdi are with Electronic and Microelectronic Laboratory, Faculty of science, University of Monastir, Tunisia (e-mail: ajmihala@gmail.com; Amira.hadjfredj@gmail.com; belgacem.hamdi@gmail.com).

Fakhreddine Zayer, Baker Mohammad, Naoufel Werghi and Jorge Dias are with the Department of Electrical and Computer Engineering, Khalifa University of Science and Technology, Abu Dhabi, UAE (e-mail: fakhreddine.zayer; baker.mohammad; naoufel.werghi; jorge.dias ; @ku.ac.ae).
Mentions of supplemental materials and animal/human rights statements can be included here.
Color versions of one or more of the figures in this article are available online at http://ieeexplore.ieee.org

accumulate (MAC), nonlinearity and compatibility with CMOS [9]–[13]. To this end, high-density memristive crossbar arrays are promising candidates for the next generation memory solutions with in-memory computing (IMC) capabilities. The later provides efficient solutions with low power consumption, low area costs and reasonable switching delays [14],[15].

However, In spite of these capabilities, NVMs suffer from a security vulnerability [16]. Since NVMs will not lose data after device power off, an attacker with physical access to the system can easily scan the memory contents and extract meaningful information from the main memory [17]. In contrast, the DRAM security relies on its short retention time for data protection [18]. To protect NVMs data, a security mechanism with a level of security comparable to that of DRAM is needed to bridge this security gap.

Real-time memory encryption (RTME), with stream cipher or pad-based, has been proposed to bridge the above security gap. In RTME, every cache memory line is encrypted or decrypted before being written and/or read [19]. The RTME is a strong protection, and it can also prevent other attacks such as memory bus snooping [20]. Unfortunately, the strong protection is at the expense of runtime performance loss, since the decryption latency, as an overhead of read memory access, is on the critical path. In addition, the access to memory encryption (ME) and memory decryption (MD) results in severe energy overhead [21]. Even though the bulk ME approach results in low performance loss at runtime, reduces the encryption tasks, and hence the energy consumption, two challenges still persist: first, it should be fast in order to lower the vulnerability window when locked and provide an instant response when unlocked. This is even more critical under the development of multicore processor and increasing demand of much larger main memory. Second, it requires energy-efficient encryption considering the limited battery life.

To address the above-mentioned challenges, we propose AES-IMC, a novel encryption architecture for fast and energy efficient NVM encryption. Embracing the benefit of the IMC architecture, AES-IMC takes advantage of multi-bit processing, large internal memory bandwidth, vast bit line-level parallelism, and low in-situ computing latency, eliminating data movement between host and memory. Leveraging the nondestructive read-out operations in NVMs for performing efficient arithmetic operations from different memristor (MR) crossbars, the entire AES procedure is performed by adding lightweight logic gates to the peripheral circuitry of the memory crossbars.

The original contributions in this paper includes:

- A new pipeline AES design. encompassing two 64-bit AES processing units instead of the conventional AES-128 unit. This approach is compatible with MR crossbar topology (i.e. 4-bit storage per cross-point element) that maintain high parallelism with the best performance, and ensures reduced area overhead under high performance and energy efficiency requirements.

- 4-bit MR-based Moore finite state machine (FSM) is proposed as the MR crossbar cross-point building element for its integration into hardware and FPGA testing. The latter supports multi-bit storage for the implementation of different arithmetic operations of AES encryption phases.

- Comparative assessment using the proposed designs with respect to the conventional AES designs and the state of the art NVM based AES implementations.

The contributions in this work do not focus on the latest advances in cryptosystem algorithms, nor on latest memory technology trends, but rather, propose a new concept of using MR designs and ICM capabilities for efficient and lightweight AES encryption, as well as their FPGA implementation. This would be of a great benefit to IoT challenges, where secure communications are of paramount importance.

The rest of the paper is organized as follow; Section II presents the background and the related works on MR integration and the AES encryption algorithm. Section III shows the pipeline and implementation of the conventional AES hardware design. Section IV describes the designs for the IMC architecture. Section V, reports the experimental results of the AES-IMC architecture in comparison with the state of the art AES architectures. Section VI. draws concluding remarks and potential future work.

## II. BACKGROUND AND RELATED WORK

There was a high interest on memristive security primitives. For instance, memristor oriented physical unclonable functions (PUFs) and true random number generators (TRNGs) were presented exclusively in [22],[23] whereas memristor oriented chaotic systems and hash functions are introduced in [24]. In general, the emerging NVM technologies are introduced in three emerging topics of hardware-oriented security where memristive solutions are expected to unfold their full potential;

i. MR cryptographic functions, e.g., block ciphers that can encrypt data and provide its confidentiality.
ii. MR entropy primitives, e.g., generation of secret keys.
iii. MR machine-learning systems, e.g., used to defend the system against attacks.

In this paper, the focus is on MR cryptographic functions for lightweight computing.

### A. MR Integration for Computing

IoT demands dedicated secure chips with constrained power consumption. The power optimization on all levels of design abstraction is the key point here, and is helpful for increasing the chip reliability and chip life. IMC is one of the leading solutions to realize area- and energy-constrained hardware systems for IoT security applications.

As shown in Fig. 2, the separation between data storage and units processing units is considered as the basic assumption for computing architectures, i.e. the von-Neumann architecture (Fig. 2a). In these system units, orders of magnitude improvement in computer performance has been achieved thanks to the impressive technological achievements. But in fact, both units in the von-Neumann architecture have reached a scaling barrier, the processing performance is now limited mostly by the so-called memory wall [25]. Various approaches have been proposed to alleviate this challenge including the integration of processing units within memory unit or the integration of cache memory near to processor [26], [27]. However, the data transfer between processing and memory units are not fully eliminated within these approaches. In the opposite, the MR trends has the potential to fully explore processing capabilities into the memory itself (Fig. 2b).

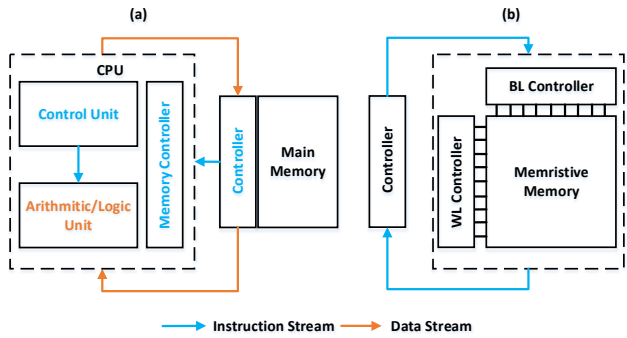

Fig. 2. General structures for (a) CMOS-based von-Neumann (b) Memristive in-memory, computing paradigms.

In-memory encryption is a promising solution for NVM encryption which has limited research. Reference [28] explores different spintronic devices-based memory that could be leveraged to implement logic functions with the AES algorithm. the efficiency of AES algorithm on a proposed in-memory processing platform with novel spin Hall effect-driven domain-wall motion devices that support both NVM cell and in-memory logic design is demonstrated in [29]. A reconfigurable cryptographic processor using in-memory computing is proposed recently [30]. By replacing a standard SRAM bank with a custom bank with in-memory and near-memory computing, Recryptor provides an IoT platform that accelerates primitive cryptographic operations. DWM is utilized to perform in-memory encryption [31], where inherent DWM device functions were used to perform the operations required for encryption.

### B. AES encryption

The AES algorithm is a symmetric block cipher capable of handling 128-bit blocks, using keys sized at 128, 192 and 256 bits. The actual key size depends on the desired security level. In this paper, AES-128 is used. This key has 128-bit full text as input and 128-bit key length. Each block of data is 4×4 array of bytes called, *State*. Due to the technology development and the high data security requirements, various approaches has been proposed in order to increase the performance of the AES algorithm. For instance, in [32], AES was implemented with the aim of reducing area and power consumption in order to maintain data throughput, achieve high speed data processing and reduce key generation time. The implemented design of AES uses pipeline structure for repeated computation by lower latency and data rates capable to support USB protocol. The design shows significant reduction in lookup tables (LUTs) and slice register compared to conventional AES. In [33], ASIC implementation of the AES based on a fully pipelined implementation, exploring the area-throughput trade-off, is presented. An achievement of 30 Gbits/s to 70 Gbits/s AES throughput is observed. In [34], a fully pipelined architecture using a high-end Virtex-7 device is proposed. 15.24 Gbits/s throughput is achieved. In [35], an efficient parallel implementation of AES on Sunway Taihulight is proposed. A high throughput of 63.91 GB/s (511.28 Gbits/s) on 1024 nodes is achieved. In [36], a hardware approach implemented on different FPGAs using block RAM resources to get an optimized architecture is proposed as well as its implementation on Artix-7 (xc7a200t) that shows a low power at the value of 0.184W. In [37], AES hardware on XILINX using Spartan-6 and Virtex-5 FPGAs is performed. The proposed design gives a higher operating frequency compared to others. In [38], a hardware design using a high-speed parallel pipelined architecture to increase the throughput to 29.77 Gbps is presented. In [39], a deep pipeline and full expansion technology to implement the AES encryption on FPGA is proposed. The architecture achieves throughput of 31.30 Gbps with a minimum latency of 0.134 us. [40] proposed a high speed implementation of AES-128 encryption with only 10 clock cycles, achieving a higher data throughput > 28 Gbps and a 220 MHz maximum frequency.

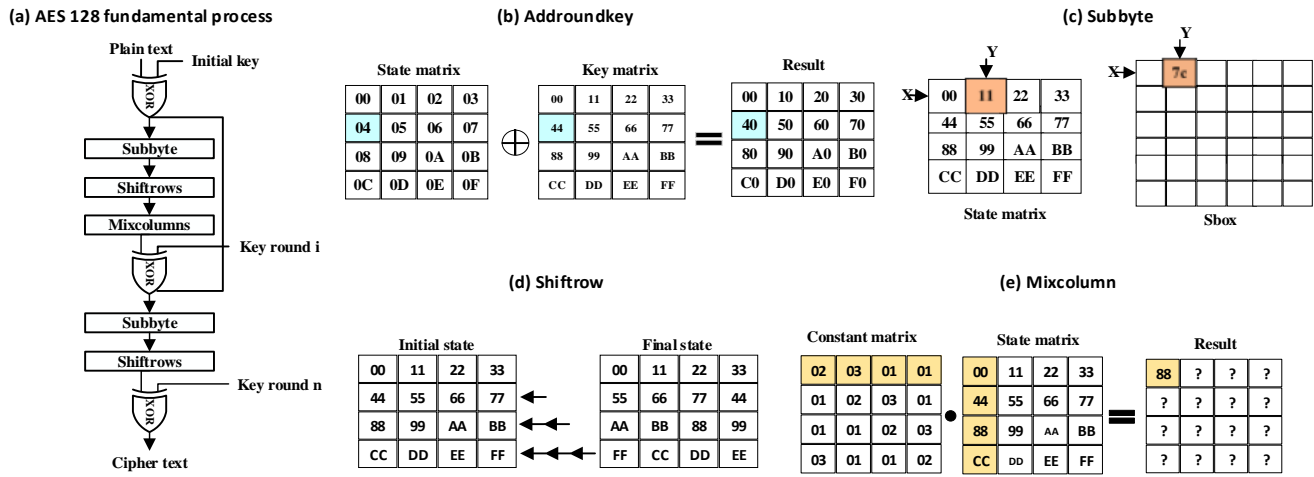

Fig.3. AES flowchart, including; (a) AES 128 process. (b) Addroundkey. (c) Subbyte. (d) Shiftrow. (e) Mixcolumn

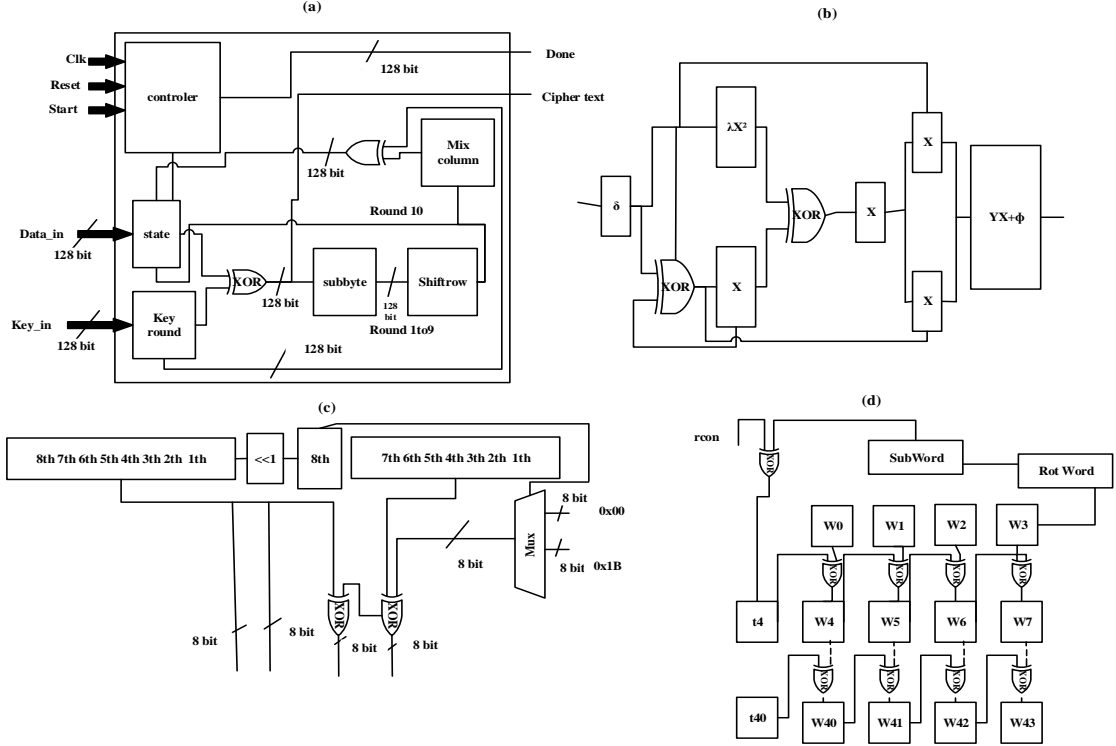

Fig. 4. Hardware logic blocks of the conventional AES encryption. (a) AES hardware design. (b) Subbyte implementation using composite field and affine transformation. (c) Output of 1 byte of Mixcolumn implementation using shifting and XOR. (d) The key generation.

In the AES, the key block round combinations with key length, $N_k$ words, and block size, $N_b$ words, in a number of rounds, $N_r$, is summarized in Fig. 3. Permutations and randomization with values of the *Subkeys* are changed at each iteration. After an initial *Addroundkey* operation, a round function consisting of four different transformations: *Subbyte*, *Shiftrow*, *Mixcolumn* and *Addroundkey* is performed. The round functions are performed iteratively 10, 12, 14 times depending on the key length. Each round needs a different round key generated by key Expansion block, which is composed by three operations: *subwords*, *rotwords* and XOR. *Subwords* and *rot-words* are only used to a specific column in the key matrix and in a specific step.

At the Addroundkey step, a XOR operation of each byte from the *state* by each byte from key matrix is needed. The *Subbyte* transformation is a non-linear byte substitution that operates independently on each byte of the *State* using the Sbox. This function can be implemented using multiplicative inversion in Galois Field (GF) ($2^8$), followed by an affine transformation, or LUTs [41]. *Shiftrow* operation is a cyclic shifting of state lines using different numbers of shift. The first row is not shifted while the second one is shifted by one byte, the third row is shifted by two bytes and the fourth is shifted by three bytes left. *Mixcolumn* transformation manipulates the data at bit level of the state bytes and multiplicated by a 4×4 constant matrix [41]. Fig. 3e) shows an example of the *Shiftrow* operation: $result[1,1] = 02.00 \oplus 03.44 \oplus 01.88 \oplus 01.CC$.

III. CONVENTIONAL HARDWARE DESIGN

In this section, a hardware block level of the AES is presented. The hardware design of the conventional AES encryption is shown in Fig. 4. This encryption process based on different transformations applied consecutively in a fixed number of iterations over the data block bits. Sub-Bytes is the costliest transformation in both time and area. The first step for finding the values of this function is to calculate the multiplicative inverse of inputs over GF($2^8$). The multiplicative inverse of $f(x)$ is $g(x)$ so that $f(x) \cdot g(x) \bmod x^8 + x^4 + x^3 + x + 1 = 1$. The second step is their computation over an affine transformation. Direct calculation of the multiplicative inverse is not easy. As all of the values are available, one of the straightforward implementations of the S-box is pre-calculating the values and storing them in a LUT read only-memory [3].

In the Mixcolumns operation, each byte is multiplied by a set of four constants ({03}, {02}, {01}, and {01}). The multiplication by 2, in GF($2^8$), can be computed by shifting the input value once to the left. If the resulting 9th bit is '1', the entire result has to be bitwise XORed (subtraction in GF($2^8$)) by '0x11B' to perform the modular reduction. The multiplication by 3 can be achieved by adding the multiplications by 1 (the input value itself) and by 2 (with the addition in GF($2^8$) being performed by a bitwise XOR.

Key Expansion algorithm is used to create the different round keys for all the 10 iterations in AES-128 and all these round keys are generated from the original (input) key, which made into 4 words $[W_0, W_3]$ where $W_0$ is the most significant word.

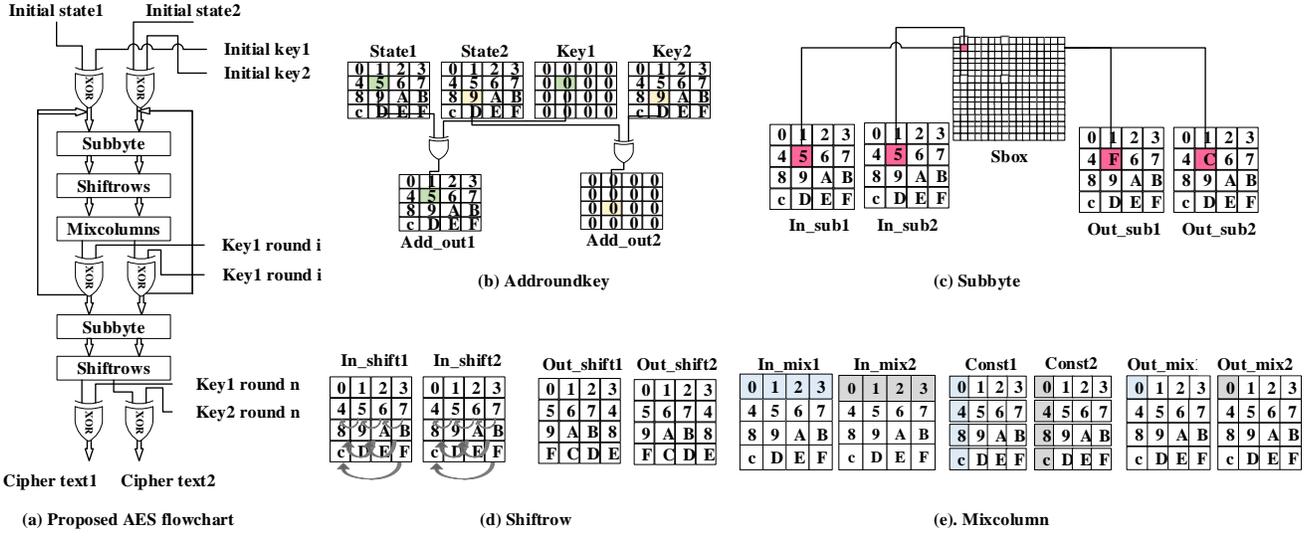

Fig. 5. Proposed AES encryption pipeline. (a) Proposed AES flowchart, (b)Addroundkey, (c) Subbyte, (d) Shiftrow, (e) Mixcolumn, procedures.

Each word is 32 bits [$W_4, W_5, ... W_{42}, and\ W_{43}$] and are the words in the remaining round keys. By doing simple Xoring between the words, this algorithm cannot produce 10 different keys. To address this, a temporary word '$t$' has to be produced for every round and given by the least significant word of previous round, which passed through nonlinear transformations Rot Word, Sub Word and an XOR operation with Round Constant *rcon* as shown in Fig. 4d). The rot word is circular (rotate) left shift to the bytes of the word, Sub word is combination of four Sub Bytes for the substitution of the bytes in the word and *rcon* is constant which is updated at each round.

## IV. PROPOSED AES IMC ARCHITECTURE

The proposed AES encryption is shown in Fig. 5. It consists of dividing the input into two sub-units. Each unit process 64-bits. All transformations of encryption process are developed following the MR crossbar purposes. The architecture is devoted to provide a high throughput and low power consumption that can be suitable for different applications. Here, we propose leveraging the multi-bit processing of memristive systems to reduce the complexity and ensure low power, latency and area performances. We employed a 4-bit MR emulator and used it as cross-point element of the MR crossbar. For each phase of the AES, different MR crossbar designs are proposed in order to imitate the arithmetic operations and functionality of the AES pipeline.

The Hardware of the proposed pipeline design is shown in Fig. 6. The scheme provides a balanced implementation with respect to low area, high throughput and low power consumption that is suitable for MR integration and useful for different applications. The plain text1 and the plain text2 are presented by input1 and input2 signals, and the encryption key1 and the encryption key2 are represented by key1 and key2 signals, respectively. The two signals input1 and key1 (input2 and key2) are XORed to perform the initial Addroundkey step of AES before the start of rounds. Input key is also applied to the key generation module, with respective rcon value, to generate the round key. The outputs are applied to an instantiated block, which is used to implement the Subbyte step of AES. The block is generated using XILINX core generator tool in which the pre-computed values stored in a ROM-based LUT. The 256 Sbox values are initialized in ROMs using *COE* coefficient file and would be wired to the ROM's address bus [42]. Shiftrow step operates on the rows of the state matrix. In this way, each column of the output state matrix of the Shiftrow step is composed of bytes from each column of the input state. This step is required to avoid the columns being encrypted independently, in which case AES degenerates into four independent block ciphers. The updated state of Shiftrow step is multiplied with the multiplication matrix to implement Mixcolumn step of AES. For multiplication, shift and XOR method are used.

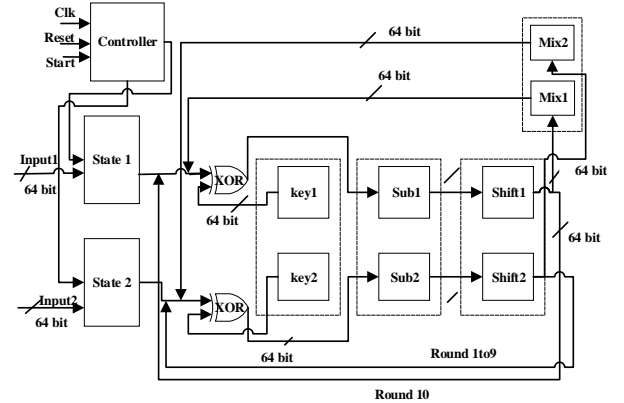

Fig 6. Hardware block level for the proposed pipeline

The AES algorithm repeats the four operations, Addroundkey, Subbyte, Shiftrow and Mixcolumn, in a certain order. The implementation detail is shown in Fig. 6. A pipelined design is used to expand the processing of the algorithm. In this way, each previous stage of the pipeline can produce all the data needed for the next step. AES requires a separate two 64-bit round key blocks for each round. Each key block can be generated by the last round key block.

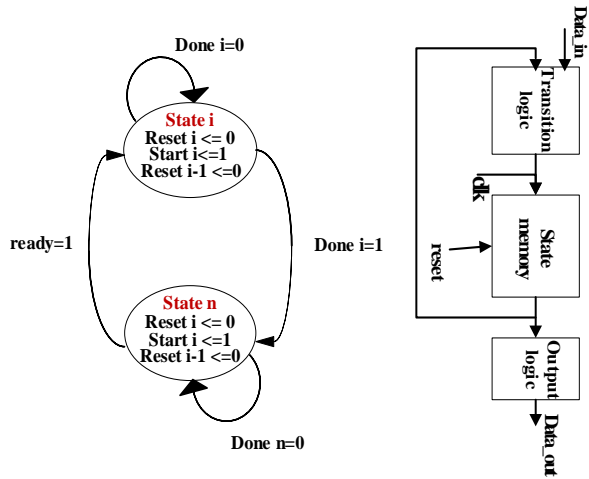

Fig. 7. Multi-bit Memristor based FSM Moore Model

*A. Memristor based Moore Finite State Machine*

Many memristive technology, nowadays supports multi-level characteristics, low latency, low power consumption [2]. The pipeline strategies and the use of IMC can be used in similar/advanced cryptographic algorithm where security and efficient low cost processing is a major concern. The MR based state machine used as the building block of the crossbar unit is presented in Fig. 7. The latter is assumed to provide 16 resistive states (4-bit).

*B. Addroundkey*

In the Addroundkey step, the main arithmetic operation is to design a XOR function within the crossbar using multi state memristor. The accumulation is implemented at the summing amplifier (SA) node. Fig. 8 shows the addroundkey transformation. First, the first row of the data matrix is read into the capacitor in each SA by activating the first word-line (red line) and selecting a column. Second, the first row of the key matrix is read into the latch in each SA by activating the second word-line (red line) and selecting a column. Then, the XOR result of two rows is latched in each SA. This Addroundkey transformation is parallelized in both write and read modes. In our design, after addroundkey transformation for one row of data, Subbyte is immediately performed for this row of data instead of continuing performing Addroundkey operation for all the four rows. The later enables high parallelism at this AES phase transition. The initial Addroundkey stage is performed with the initial key. The other 10 rounds are performed with the corresponding round keys. The key generator shared among all memory banks generates the key and sends it to different banks. After each round of encryption, the key generator expands the initial key to get the corresponding round key. As shown in Fig.3, the encryption key is maintained in the MR memory array and round keys overwrite the encryption key after each round of encryption.

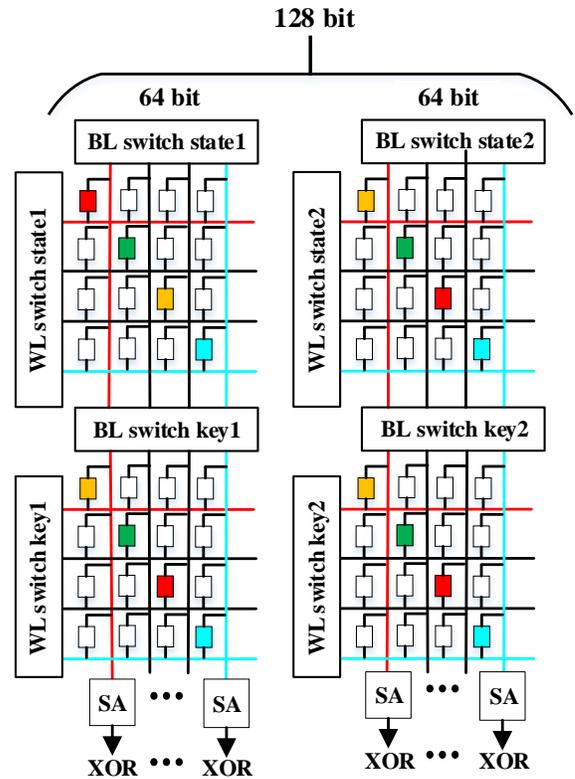

Fig. 8. Addroundkey AES-IMC based XOR design.

*C. Subbyte*

For Subbyte transformation, each 4bit of the two data matrix is decoded and inputted to the S-box as shown in Fig. 9. The Addroundkey results of the second row of data matrix are latched in the SAs. Subbyte is performed on the second cell. The output of S-box is the substituted byte. The S-box combinational logic has 8-bit input and 8-bit output. Since we can only input one byte each time to the S-box, the Subbyte transformation can only be done sequentially which takes a long time. To accelerate the Subbyte transformation, more Sbox combinational logic are made available for parallel Subbyte operation. At the same time, we need to consider the hardware overhead introduced by multiple S-box.

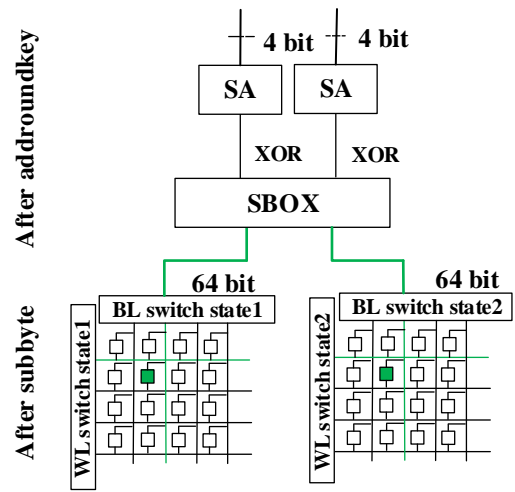

Fig.9. Subbyte AES-IMC design

### D. Shiftrow

Shiftrow transformation is realized with address decoding and control signals. The 4-bit output of S-box of each matrix needs to be written back where each input bit is located. By combining an offset with the column address, the S-box output is shifted to another address w.r.t, to the Shiftrow topology presented in Fig.10. This shifting process is done by selecting the first column along with the control signal. After Shiftrow transformation, each bit will be buffered in the single-bit latch until Subbyte and Shiftrow are performed in the SAs. The values in the row buffer are then transmitted to the write driver and written back to the memory array. This row buffer holds the intermediate results and avoids extra writes to the non-volatile memory rows.

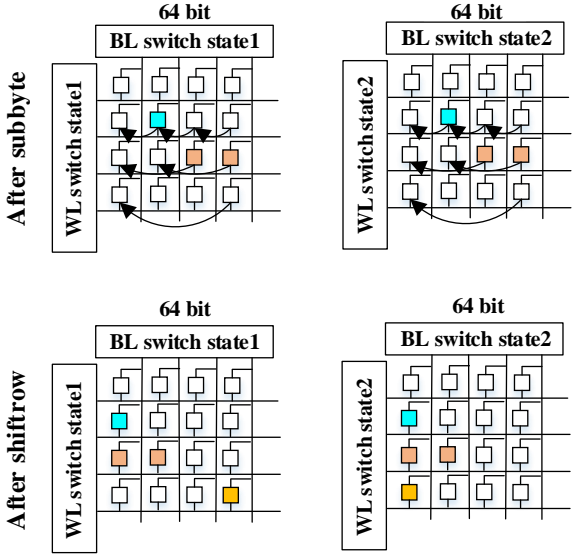

Fig.10. Shiftrow AES-IMC design

### E. Mixcolumn

In the Mixcolumns stage, the data matrix is multiplied by a known matrix. In this way, each column of the data matrix is combined with the column of key using an invertible linear transformation to provide diffusion in the cipher results as shown in Fig. 11. This matrix multiplication is done in the finite field $GF(2^8)$, which can be decomposed to modular multiplication and XOR operations.

We use $S_{i,j}$ and $S'_{i,j}$ to indicate the byte in row i, column j of the state matrix and the transformed state matrix respectively. The Mixcolumns transformation is as follows:

$$\begin{aligned} S'_{0,j} &= 2 \cdot S_{0,j} \oplus 3 \cdot S_{1,j} \oplus S_{2,j} \oplus S_{3,j} \\ S'_{1,j} &= S_{0,j} \oplus 2 \cdot S_{1,j} \oplus 3 \cdot S_{2,j} \oplus S_{3,j} \\ S'_{2,j} &= S_{0,j} \oplus S_{1,j} \oplus 2 \cdot S_{2,j} \oplus 3 \cdot S_{3,j} \\ S'_{3,j} &= 3 \cdot S_{0,j} \oplus S_{1,j} \oplus S_{2,j} \oplus 2 \cdot S_{3,j} \end{aligned} \quad (1)$$

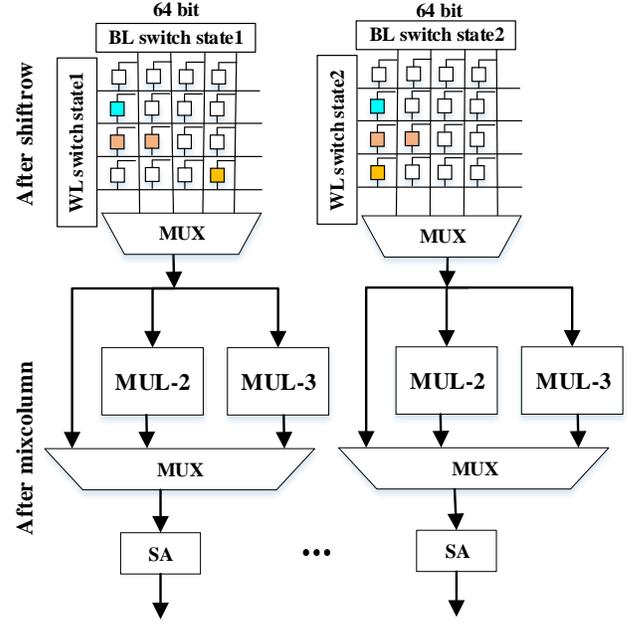

Fig.11. Mixcolumn AES-IMC design

Multiplication-by-2 (M-2) in the finite field can be realized by leveraging an LUT. Mixcolumn can be realized with M-2 and XOR logic.

$$3 \cdot S_{i,j} = 2 \cdot S_{i,j} \oplus S_{i,j} \quad (2)$$

Therefore, Mixcolumns stage is decomposed into M-2 LUT and XOR operations. Mixcolumns generates several intermediate values. To both accelerate this transformation and maintain a low hardware overhead, we leverage the vacant non-volatile memory rows as buffer rows for intermediate results. The MixColumns stage is realized with LUT and XORs as follows:

$$\begin{aligned} S'_{0,j} &= Tj \oplus 2 \cdot S_{0,j} \oplus 2 \cdot S_{1,j} \oplus S_{0,j} \\ S'_{1,j} &= Tj \oplus 2 \cdot S_{1,j} \oplus 2 \cdot S_{2,j} \oplus S_{1,j} \\ S'_{2,j} &= Tj \oplus 2 \cdot S_{2,j} \oplus 2 \cdot S_{3,j} \oplus S_{2,j} \\ S'_{3,j} &= Tj \oplus 2 \cdot S_{0,j} \oplus 2 \cdot S_{3,j} \oplus S_{3,j} \end{aligned} \quad (3)$$

*where*

$$Tj = S_{0,j} \oplus S_{1,j} \oplus S_{2,j} \oplus S_{3,j} \quad (4)$$

The first step of Mixcolumns is the M-2 transformation. This process shares the same address decoding logic of S-box with a MUX. Since only one byte can be input each time to the LUT, this transformation is done sequentially. To accelerate this process, multiple M-2 LUTs are added to enable parallel operations. Like the S-box design, different multiplication-by-2 LUT designs are considered with different encryption speed and overhead. After M-2 transformation, outputs are latched in a row buffer until all bytes of the activated row finishes M-2 transformation. Then data in this row buffer is written to a vacant memory row. The next step of Mixcolumn is calculating $Tj$ following (4). Every time two rows are activated to get the XOR result of two memory cells, this result is written to an empty buffer row. In this step, since all SAs are working simultaneously, $Tj$ for each column is calculated in full parallel. The final step of Mixcolumns is calculating the result of Mixcolumns transformation as described in (3). In this step, with M-2 LUT results stored in four rows and $Tj$ values, Mixcolumn for one row of selected columns can be completed in six steps. Four operands are XORed together to get S′0,j, the result is then written back replacing $S_{0,j}$.

## V. Experimental Results and Evaluation

In this section, we present and discuss the results of the proposed AES-IMC design with respect to other AES hardware and memristive based architectures, following a set of metrics. The proposed AES-IMC is implemented on the NEXYS 4 DDR, Artix-7, FPGA board. The synthesis process for the designs was configured with *Area* as *Optimization Goal* and *High* as *Optimization Effort*. The FPGA implementation and the synthetized design after simulation are shown in Fig. 12. The input size is 128 bits divided in two parts, each one of 64 bits, verifying that the complete encryption of the data packet is done on 26 cycles. Fig.13 shows a simulation of the conventional AES and the proposed AES-IMC designs. The source files for the proposed architecture were provided by the authors on https://github.com/FakhreddineZ/Hardware-Security.

### A. Metrics for Evaluation

The set of metrics used for evaluating the AES-IMC hardware design are area, latency exhibited and the throughput achieved by the hardware and energy consumption. Slices (SLC) are used as area units, but the results in LUT and Flip-Flops (FF), usage are also presented and discussed. In terms of energy, the calculation of the total power dissipated by the architectures and the energy required to process a plaintext block are provided.

The throughput-per-slice, $Thr/SLC$, and the energy-per-bit E/bit, are used as derived metrics to evaluate the efficiency of the proposed architecture. The former is a relation of the implementation size and latency. The latter is the estimated energy required to process the state.

The maximum throughput, $Thr$ of an implementation is a function of the maximum operating frequency, $F_{max}$, the latency of clock cycles consumed in the longest stage of AES implementation, $L$, and the block size $B_{size}$, as per the following formula:

$$Thr = \frac{F_{max} \times B_{size}}{L} \quad (5)$$

The throughput per slice $Thr/SLC$ is simply calculated using this ratio:

$$Thr/SLC = \frac{Thr}{SLC} \quad (6)$$

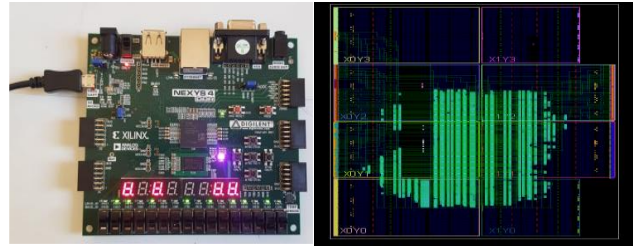

Fig.12. a) FPGA Implementation, b) Floor plan of the AES-IMC synthesis.

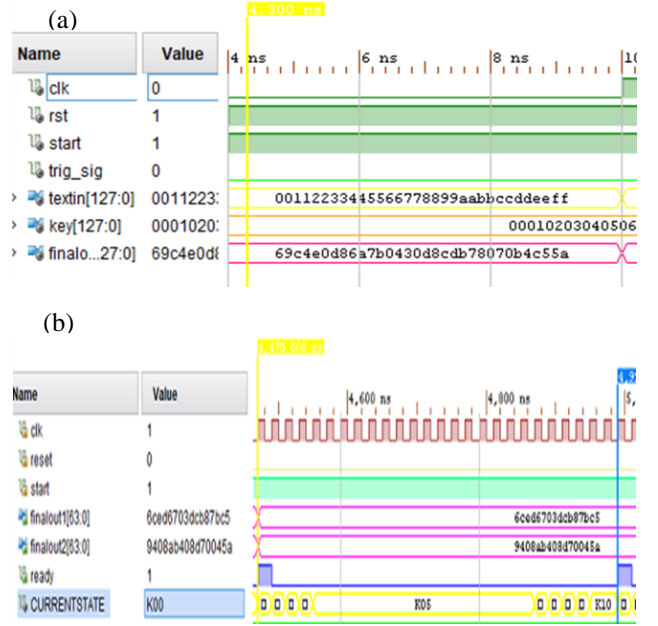

Fig. 13. Simulation, a) Conventional AES output. b) Proposed AES-IMC output.

Regarding performance, the maximum operational frequency of each architecture, and two constant operational frequencies of 13.56 MHz and 30Mhz across are used for fair comparison. The former can be of use to systems that require low resource usage but also have high performance constraints determined by the application. The later enables a comparison across all the implementations which can be useful for systems that require low resource usage and can accept compromises in the performance. The frequency of 13.56 MHz is utilized since it is appropriate for RF applications, which is the case of some IoT transmitters. A clock-rate of 30MHz is used for comparing the AES accelerators for achieving an optimal energy efficiency [43]. Operating the accelerators at faster clock-rate will adversely affect the systems reliability due to high peak power for densely distributed and concurrent execution. Due to the number of concurrent encryptions within a small footprint, 30MHz is a reasonable clock rate according to all custom circuit implementations to avoid overheating [31]. These metrics will be used to discuss the performance results.

TABLE I.
RESOURCE USAGE AND PERFORMANCE RESULT FOR DIFFERENT ARCHITECTURES

| Work | Device | State (bit) | Key (bit) | FF | LUT | SLC | FMAX (MHz) | L (cycles) | Thr (Mbps) | Thr/SLC (Mbps) | Thr* (Mbps) |
|---|---|---|---|---|---|---|---|---|---|---|---|
| [47] | XC6SLX16-3CSG324 | 64 | 128 | 200 | 202 | 58 | 160.21 | 55 | 186.42 | 3.214 | 15.78 |
| [48] |  | 64 | 128 | 73 | 147 | 48 | 206.40 | 132 | 100.07 | 2.085 | 6.57 |
| [49] |  | 64 | 128 | 201 | 220 | 61 | 210.66 | 136 | 99.13 | 1.625 | 6.38 |
| [47] | XC3S200-5FT256 | 64 | 128 | 200 | 381 | 191 | 179.95 | 55 | 209.40 | 1.096 | 15.78 |
| [48] |  | 64 | 128 | 73 | 280 | 153 | 120.71 | 132 | 58.53 | 0.383 | 6.57 |
| [49] |  | 64 | 128 | 201 | 264 | 151 | 194.63 | 136 | 91.59 | 0.607 | 6.38 |
| [47] | XC5VLX50T-3FF1136 | 64 | 128 | 200 | 283 | 88 | 271.67 | 55 | 316.12 | 3.592 | 15.78 |
| [48] |  | 64 | 128 | 73 | 182 | 75 | 321.96 | 132 | 156.10 | 2.081 | 6.57 |
| [49] |  | 64 | 128 | 201 | 239 | 73 | 431.78 | 136 | 203.19 | 2.783 | 6.38 |
| [47] | XC4VLX25-12FF668 | 64 | 128 | 200 | 382 | 192 | 284.33 | 55 | 330.86 | 1.723 | 15.78 |
| [48] |  | 64 | 128 | 73 | 279 | 151 | 223.51 | 132 | 108.37 | 0.718 | 6.57 |
| [49] |  | 64 | 128 | 201 | 265 | 152 | 364.56 | 136 | 171.56 | 1.129 | 6.38 |
| [36] | Artix7 XC7A200T | 128 | 128 | 2911 | 1512 | 359 | 311.72 | 59 | 676.276 | 1.884 | 29.41 |
| [47] | Virtex II | 128 | 128 | 271 | 1862 | 976 | 60.94 | **26** | 300.01 | 0.307 | **66.76** |
| [50] | Virtex IV XC4VLX200 | 128 | 128 | 684 | 2127 | 1120 | 112.37 | 1000 | 14.383 | 0.013 | 1.73 |
| [47] | Virtex V | 128 | 128 | 271 | 1391 | 456 | 96.04 | **26** | 472.81 | 1.037 | **66.76** |
| [36] | Virtex7 XC7VX90T | 128 | 128 | 1817 | 1271 | 551 | 308.64 | 59 | 669.59 | 1.215 | 29.41 |
| [40] | XCZU9EG | 128 | 128 | 4296 | 15029 | 3262 | 220 | **10** | 2816 | 0.863 | **173.56** |
| [51] | Spartan 7 XC7S75FGGA484-1 | 128 | 128 | 1900 | 1425 | 430 | 213.7 | 57 | 480 | 1.116 | 30.45 |
| AES-IMC | Xc7A100T-CSG324 | 128 | 128 | 1245 | 2836 | 468 | 108.9 | **26** | 536.12 | **1.144** | **66.76** |

TABLE II.
POWER AND ENERGY CONSUMPTION RESULTS FOR DIFFERENT FPGA IMPLEMENTATIONS

| Work | Device | State (bit) | Key (bit) | L (cycles) | P (mW) | E*(μJ) | E*/Bit(nJ) |
|---|---|---|---|---|---|---|---|
| [47] | Xc6slx16-3csg324 | 64 | 128 | 55 | 21.31 | 0.086 | 1.351 |
| [48] |  | 64 | 128 | 132 | 21.6 | 0.210 | 3.285 |
| [49] |  | 64 | 128 | 136 | 21.76 | 0.218 | 3.410 |
| [47] | Xc3s200-5ft256 | 64 | 128 | 55 | 41.97 | 0.170 | 2.660 |
| [48] |  | 64 | 128 | 132 | 42.37 | 0.412 | 6.445 |
| [49] |  | 64 | 128 | 136 | 42.36 | 0.425 | 6.638 |
| [47] | Xc5vlx50t-3ff1136 | 64 | 128 | 55 | 563.51 | 2.286 | 35.71 |
| [48] |  | 64 | 128 | 132 | 563.57 | 5.486 | 85.720 |
| [49] |  | 64 | 128 | 136 | 562.67 | 5.643 | 88.177 |
| [47] | Xc4vlx25-12ff668 | 64 | 128 | 55 | 348.88 | 1.415 | 22.11 |
| [48] |  | 64 | 128 | 132 | 242.38 | 2.359 | 36.866 |
| [49] |  | 64 | 128 | 136 | 248.02 | 2.488 | 38.867 |
| [36] | Artix7(XC7A200T) | 128 | 128 | 59 | 0.184 | 0.80 | 6.25 |
| [50] | Virtex IV XC4VLX200 | 128 | 128 | 1000 | 0.261 | 19.247 | 150.37 |
| [36] | Virtex7(XC7VX90T) | 128 | 128 | 59 | 0.463 | 2.01 | 15.7 |
| [40] | XCZU9EG | 128 | 128 | 10 | 1.17 | 0.86 | 6.74 |
| [51] | Spartan 7 xc7s75fgga484-1 | 128 | 128 | 57 | 563.5 | 2.368 | 18.5 |
| AES-IMC | Xc7A100T-CSG324 | 128 | 128 | **26** | **0.098** | **0.18** | **1.406** |

However, low-resource implementations usually limit the spectrum of used frequencies to lower frequencies, as is the case in RFID applications, i.e. $F_{RF}$=13.56MHz [44]. By having a common frequency, to evaluate the proposed architecture with other AES FPGA implementation, a fair comparison in terms of $Thr/SLC$ can be performed. The throughput calculated at $F_{RF}$ is denoted $Thr^*$ and calculated as follows:.

$$Thr^* = \frac{F_{RF} \times B_{size}}{L} \qquad (7)$$

The energy, $E$, expended by the implementation to process a single block is calculated as below, where $P$ is the total dissipated power.

$$E^* = \frac{P \times L}{F_{RF}} \qquad (8)$$

(8) is used to estimate the energy-per-bit.

TABLE III.
PERFORMANCE COMPARISON OF DIFFERENT AES-128 bit TECHNOLOGICAL DESIGNS

| Work | Area (μm²) | L(cycles) | P (W) | FMAX (MHz) | E (nJ) | Thr (Mbps) |
|---|---|---|---|---|---|---|
| CMOS ASIC [45] | 4400 | 336 | 0.013 | 30 | 6.6 | 5.16 |
| Memristive CMOL [46] | 320 | 470 | 0.309 | | 10.3 | 3.69 |
| Baseline DW-AES [52] | 78 | 1022 | 0.072 | | 2.4 | 1.71 |
| Pipelined DW-AES [52] | 83 | 2652 | 0.069 | | 2.3 | 0.65 |
| Multi-issue DW-AES [52] | 155 | 1320 | 0.081 | | 2.7 | 1.31 |
| AES-IMC | ~83 | **26** | 0.098 | | **0.9** | **147.6** |

TABLE IV.
SYSTEM CONFIGURATION: AES COMPUTING PLATFORM CONFIGURATIONS AND DPR UNDER 2MM² AREA BUDGET

| Platform | $\#C_i$ | L(128bits) | *DPR* (GB/s) |
|---|---|---|---|
| CMOS ASIC [45] | 454 | 84 | 2.59 |
| Memristive CMOL [46] | 6250 | 470 | 6.38 |
| DW-AES Baseline [52] | 2564 | 1022 | 12 |
| DW-AES Pipelined [52] | 24096 | 663 | 17.4 |
| DW-AES Multi issue [52] | 12902 | 220 | 28 |
| AES-IMC | ~24096 | 26 | **445** |

$$E^*/bit = E^*/B_{size} \quad (9)$$

In addition, given a 30Mhz uniform operating frequency, the number of cycles for the most time-critical stage is different among all implementations, which is indicated in Table IV for a 128-bit cipher. As figure of merit, the data processing rate (DPR) is defined to measure the rate of encrypted data within a given area budget for different custom implementations. The DPR is defined as

$$DPR = \frac{\#C_i \times F_{RF} \times \#B_{C_i}}{L} \quad (10)$$

where $\#C_i$ is the number of ciphers that can be implemented in a given are, $\#B_{C_i}$ is the number of bytes in each cipher and the latency of critical stage is the number of clock cycles consumed in the longest stage of AES implementation.

*B. Area*

Table I presents the performance and resource usage results for the different architectures under evaluation in the different FPGAs devices. The resource usage for all the architectures is presented for the diffetrent FPGAs selected as implementation platform, this metric is illustrated in Fig. 14. The results are consistent for LUT-4 FPGAs. In the case of the LUT-6 platforms, for instance, it can be noted how implementations in the Spartan-6 FPGA use less LUT elements, which derives in lower slice counts than those of the implementations in the Virtex FPGAs and our implementation. This is due to slight variations in the slice architecture for both FPGAs yielding different results depending on the implementation strategies. The proposed architecture achieves comparable resource usage than the compared iterative designs when implemented in LUT-4 FPGAs.

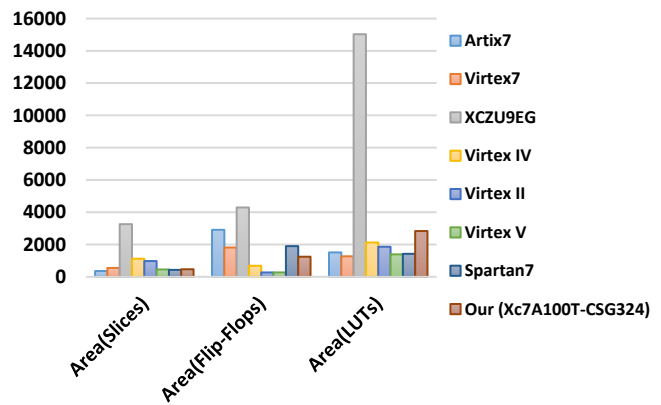

Fig. 14. Slices, Flip-Flops, LUTs resource usage for the different 128 State architectures in the different FPGAs utilized.

Benefiting from the high density of multi-bit processing in memory crossbar, the proposed architecture is highly area-efficient. In the following, the hardware resources (HR) of our proposed architecture, is concluded based on the number of LUT, FF, and input/output (I/O). Given a similar approach of using in-memory computing for AES [31], The areas of ShiftRows, AddRoundKey and MixColumns are estimated as 0.04μm², 3.3μm² and 0.3μm², respectively. Additional registers and I/O for pipeline and hardware resources. Overall, the DW-AES cipher saves 98% and 76% of areas over the CMOS ASIC [45] and memristive CMOL [46] designs, respectively. Compared to DW-AES [31], the pipelined DW-AES incurs an area overhead and that is due mainly to the single bit processing and DW-FIFO used for stage balancing and additional state matrices. The breakdown of areas consumed by different modules of the implementations is shown in Fig. 15. Due to the use of state machine emulator for 4bit memristor, I/Os in all AES phases consume almost of the area.

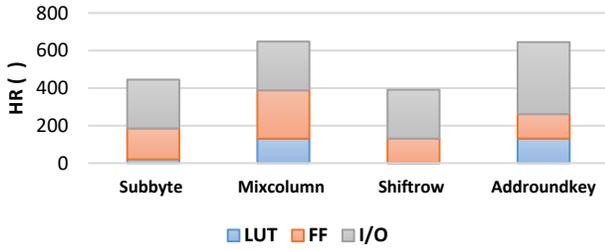

Fig. 15. Hardware resources for different phases of the AES-IMC architecture.

### C. Latency and Throughput

The premium for the lower area and energy efficiency of AES-IMC over the other technological implementation is the decrease of the number of cycles required for its computation. The number of cycles for different FPGA implementations is presented in table I. the multi bit processing in memory crossbar avoid the additional cycles required in DW-FIFO, for instance to balance the number of cycles of all pipelined stages and the use of multiple processing units due to single bit processing. In addition, the latency of the pipelined design is boosted by a higher processing rate as it will be shown later.

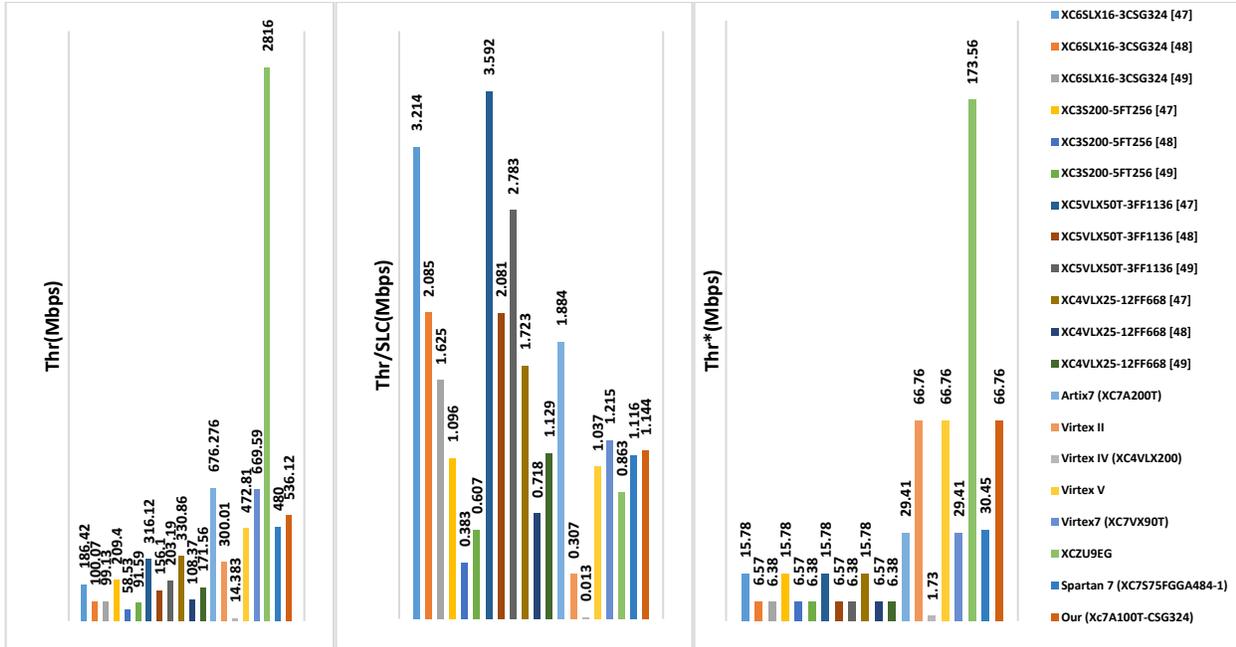

Fig. 16. Throughput for the different designs, using different FPGAs and different operational frequencies. (a) Throughput using the maximum frequency. (b) Throughput/Slice using the maximum frequency. c) Throughput at 13.56 MHz.

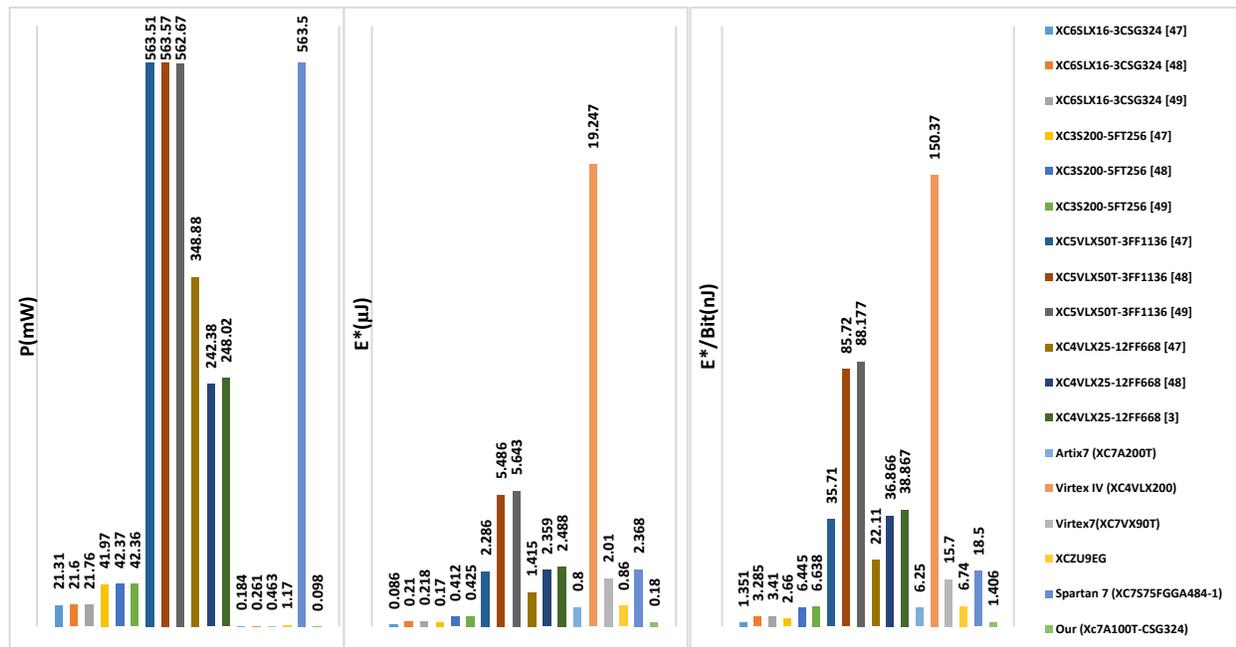

Fig. 17. Total power and energy analysis for the different configurations in the different FPGAs utilized. (a) Total power, b) Energy consumption. (b) Energy-per-bit.

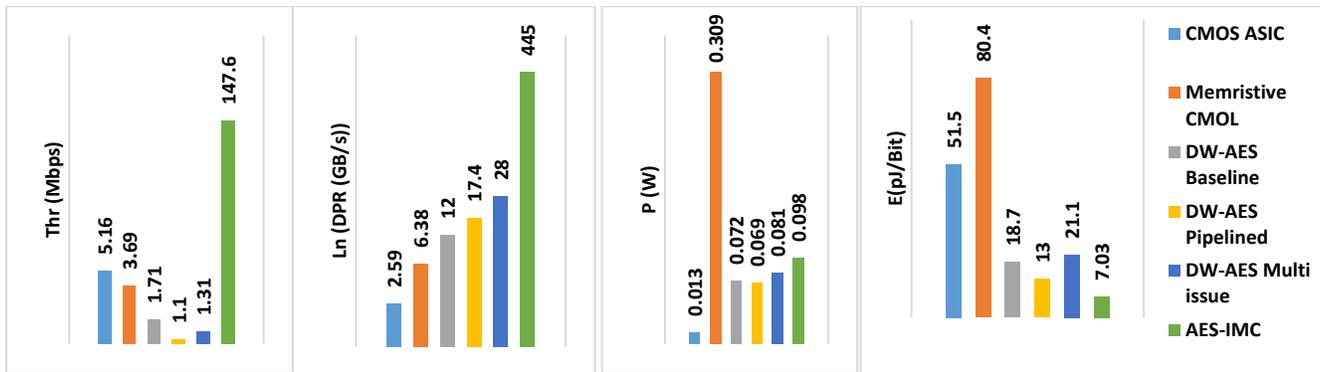

Fig. 18. Comparison of throughput, processing rate (DPR), power and energy efficiency among different AES-IMC computing platforms.

As for he throughput, it is calculated using a constant frequency to be matched with the energy consumption analysis, which is calculated using the same frequency. Our AES-IMC implementation provides a maximum frequency to 108.9MHz. From the maximum frequency it can be noted that the results depend not only on the implementation, but also on the underlying FPGA platform. The performance comparison for the different implementations is shown in Fig. 16 using the maximum frequency of the implementation and frequency recommended in [44] of 13.56MHz, The throughput per slice is a metric utilized to illustrate the efficiency of the architectures when it is desired to study the trade-off between the area reduction and the performance of the implementations. In this case the non-optimized implementations will have the maximum efficiency, and the area-optimized architectures can be ranked from this reference. Note that the implementation with a reduced key size should be compared having in mind that it also features a security trade-off.

### D. Power, Energy Efficiency and Data Processing Rate Comparison

Table II presents the power and energy consumption results using different FPGA platforms. Their corresponding graphic comparison for the different implementations is shown in Fig. 17. The analysis performed for each architecture delivers power and temperature estimations based on a user defined operational frequency and temperature. It was determined to use a frequency of 13.56 MHz for all the studies and the default operation temperature. Since the goal of this experiment is to study the energy consumption, only the power results were used. In most of them it is shown how the static power remains constant across the different implementations for the same FPGA board. Regarding the dynamic power, it can be noted how it changes depending on the switching activity of the circuit. The total power is the sum of the static and dynamic power. The power analysis demonstrates how selecting the appropriate FPGA board can deliver a change with a significance of an order of magnitude. The energy per bit metric, also as an efficiency measurement, represents the energy cost associated to process a single bit of the plaintext. This can be of interest to compare these results with those obtained from implementations of algorithms with different state size. In Fig. 17c) the energy per bit for the different implementations is presented.

In addition, the proposed AES-IMC is compared with other recent technology based implementations at the system level. For each AES computing platform, the number of AES units is maximized to encrypt the input data stream concurrently subject to a fixed area constraint. Given a 2mm$^2$ area design budget, the system configurations for different platforms are summarized in Table III and Table IV. Fig. 18 compares throughput, DPR, power and energy efficiency of different AES computing platforms. Among all the hardware implementations, the proposed AES-IMC has a DPR of 445 GB/s, which is 171.8× higher than that of the CMOS ASIC based platform with a 86.3% energy E/Bit reduction., and 69.7× higher than that of the memristive CMOL based platform with 91.2% energy E/Bit reduction. The proposed architecture presents the highest throuhout over these technologies based platforms. Due to the smaller area per cipher using the AES-IMC, more ciphers can operate in parallel under the same area budget, leading to its higher DPR. This further improvement in DPR is achieved by reducing the latency of the critical stage. Due to the multi bit in-memory encryption and non-volatility, the proposed AES-IMC computing platforms have the best energy efficiency of 7.03pJ/bit, which is higher than that of state of the art platforms. The advantage of energy efficiency of the proposed architecture over ASIC implementations remains competitively better owing to its efficient in-memory communication and lower energy consumption per encryption.

### VI. CONCLUSION

The paper presents an efficient tool for AES encryption. Techniques for increasing the efficiency of AES processing and MR design exploration are proposed. The proposed AES-IMC encrypts many memory blocks simultaneously within the entire encryption process and completed within the main memory without exposing the results to the memory bus. Experimental results show that AES-IMC outperforms state-of-the-art AES engines with higher throughput and higher data processing rate yet lower energy consumption. In the future work, along with the current AES-IMC, we aim to enhance the security level of the system by in introducing memristive chaos and its random key generation toward efficient computing and high secure system. This encryption architecture will prevent unintended accidents with unmanned devices, also caused by malicious attacks. This research can increase the cyber security of autonomous driverless cars or robotic autonomous vehicles.